\documentclass{chi-ext}

\copyrightinfo{\scriptsize Permission to make digital or hard copies of part or all of this work for personal or classroom use is granted without fee provided that copies are not made or distributed for profit or commercial advantage and that copies bear this notice and the full citation on the first page. Copyrights for third-party components of this work must be honored. For all other uses, contact the owner/author(s). Copyright is held by the author/owner(s). \\
{\emph{CHI'15 Extended Abstracts}}, April 18--23, 2015, Seoul, Republic of Korea. \\
ACM 978-1-4503-3146-3/15/04. \\
http://dx.doi.org/10.1145/2702613.2732781}
\title{Ariadne's Thread --- Interactive Navigation in a World of Networked Information}

\numberofauthors{4}

\author{
  \alignauthor{
  	\textbf{Rob Koopman}\\
  	\affaddr{OCLC \\
    Schipholweg 99, \\Leiden, The Netherlands}\\
  	\email{rob.koopman@oclc.org}
  }
  \vfil
  \alignauthor{
  	\textbf{Shenghui Wang}\\
  	\affaddr{OCLC }\\
  	\affaddr{Schipholweg 99}\\
  	\affaddr{Leiden, The Netherlands}\\
  	\email{shenghui.wang@oclc.org}
  }
  \vfil
  \alignauthor{
  	\textbf{Andrea Scharnhorst}\\
  	\affaddr{Royal Netherlands Academy of Arts and Sciences}\\
   	\email{andrea.scharnhorst@dans.knaw.nl}
  }
  \vfil
  \alignauthor{
  	\textbf{Gwenn Englebienne}\\
  	\affaddr{University of Amsterdam}\\
  	\affaddr{Science Park 904}\\
  	\affaddr{Amsterdam, The Netherlands}\\
  	\email{G.Englebienne@uva.nl}
  }
}

\def\plaintitle{}
\def\plainauthor{}
\def\plainkeywords{}
\def\plaingeneralterms{}

\hypersetup{
  pdftitle={\plaintitle},
  pdfauthor={\plainauthor},  
  pdfkeywords={\plainkeywords},
  pdfsubject={\plaingeneralterms},
}

\usepackage{graphicx}   
\usepackage{balance}    
\usepackage{bibspacing} 
\usepackage[usenames,dvipsnames,svgnames,table]{xcolor}
\usepackage{xspace}
\usepackage{enumitem}
\setlist{noitemsep,topsep=0pt,parsep=0pt,partopsep=0pt}

\newcommand{\fig}[1]{Figure~\ref{fig.#1}}

\begin{document}

\maketitle

\begin{abstract}
This work-in-progress paper introduces an interface for the interactive visual exploration of the context of queries using the ArticleFirst database, a product of OCLC. We describe a workflow which allows the user to browse live entities associated with 65 million articles. In the on-line interface, each query leads to a specific network representation of the most prevailing entities: topics (words), authors, journals and Dewey decimal classes linked to the set of terms in the query. This network represents the context of a query. Each of the network nodes is clickable: by clicking through, a user traverses a large space of articles along dimensions of authors, journals, Dewey classes and words simultaneously. We present different use cases of such an interface. This paper provides a link between the quest for maps of science and on-going debates in HCI about the use of interactive information visualisation to empower users in their search.  
\end{abstract}

\keywords{\plainkeywords}
knowledge maps; interfaces to digital libraries; science maps; random projection; interactive visualisation

\category{H.5.2}{Information Interfaces and Presentation}{User Interfaces}
\category{H.3.3}{Information Storage and Retrieval}{}. 

\section{Motivation}
With digitization and the world wide web, any information seems to be at our fingertips. And yet it is very cumbersome to investigate a topic, to understand its context and history, and to find authoritative sources for it. It has been stated that information retrieval based on pure text-statistical methods has reached a certain limit~\cite{Mutschke2014}. For current information retrieval, it is difficult to answer questions such as:
\begin{itemize}[noitemsep,topsep=0pt,parsep=0pt,partopsep=0pt,leftmargin=*]
\item What are the different aspects of this topic?
\item Are there related aspects missing in my search terms? 
\item Who are the most prominent authors about this topic?
\item Which journals publish most about this topic?
\item How have others --- e.g. librarians --- described and classified this topic?
\end{itemize}
For scientific articles, the entities involved in these questions---authors, journals, subjects---are essentially interlinked and can be accessed via bibliographic databases such as the \textit{Web of Science},
\textit{Scopus}, \textit{SpringerLink}, \textit{ArticleFirst} or \textit{Microsoft Academic Search}. Back in the history of information science, the \textit{Web of Science}
has fostered the emergence of a whole new field: scientometrics. With the introduction of the principle of citation indexing \cite{garfield55} and access to networks of scientific papers, there have been dreams to visualize this fabric of science \cite{waltman2010}. The making of science maps has culminated in the exhibition \textit{Places\&Spaces} \cite{boerner2010atlas}. It also resulted in discussions on how to implement them into digital libraries \cite{boernerchen2002interfaces}. To visualize the context of an scholarly argument has further inspired the maker of interactive interfaces \cite{PivotPaths,citeology}  Still, bibliographic databases are dominated by a single search term window and ranked lists of results. Maps are only occasionally implemented. Examples are the AuthorMapper\footnote{\small\url{http://www.authormapper.com/}} for \textit{SpringerLink}, the co-author graph for \textit{Microsoft Academic Search},\footnote{\url{http://academic.research.microsoft.com/VisualExplorer\#899177}} or \textit{HistCite} for \textit{Web of Science}.\footnote{\url{http://garfield.library.upenn.edu/histcomp/price-djd_auth/graph/1.html}} Most of the time, visual exploration is possible in stand-alone tools such as the \textit{Sci\textsuperscript{2}} tool\footnote{\url{https://sci2.cns.iu.edu/user/index.php}} or \textit{CitNetExplorer}.\footnote{\url{http://www.citnetexplorer.nl/Home}} Visual navigation has been also implemented for specific projects such as the \textit{MESUR} project for clickstream data\footnote{\url{http://mesur.informatics.indiana.edu/cgi-bin/demos/map.cgi}} or the \textit{GenderBrowser}\footnote{\url{http://www.eigenfactor.org/gender/}} which operates on a dump of JSTOR. At the same time, information and computer scientists (HCI, InfVis, Visual Analytics) call for further experiments to navigate visually and interactively through large information spaces~\cite{dork2012,akdag2012,shneiderman2000axes}. This paper contribute to further connect different discourses about analysing, visualizing and navigating large bodies of scholarly knowledge.

Obviously, it would be valuable if in the search for topics we could seamlessly travel between relevant subjects, authors and journals. Imagine, starting with a search term, we could find other subjects related to it; after choosing a subject, the most relevant authors or journals could be presented; for an author, we would be able to find subjects he/she publishes about and which journals are most relevant to those. In such an envisioned scenario the user can shift his/her interests as the exploration goes on. \textit{Serendipity} is not only expected, but full-heartily embraced. In this work-in-progress paper, we present an interface which enables such journeys. To be able to \textit{scroll through information spaces} has motivated many explorations of interactive interfaces~\cite{PivotPaths}. This paper joins into those efforts but takes a slightly different approach, focussing on similarities in large vector spaces composed by terms, author, journal and Dewey classes.

\paragraph{Dataset}
We develop a visual interactive interface to browse through contexts based on information from the article database \textit{ArticleFirst} of OCLC. 
ArticleFirst contains more than 65 million article records from more than 30 thousand journals with which more than 3 million authors are associated. We treat author names, journal ISSNs, Dewey classes (assigned to journals) and topical terms (words) in the title and abstract of articles as entities. The method allows us to investigate the relations between roughly 1 million topical terms, 3 million authors, 30 thousand journals and 738 Dewey decimal classes.

\section{Method}

\marginpar{
\begin{figure}\includegraphics[width=4cm]{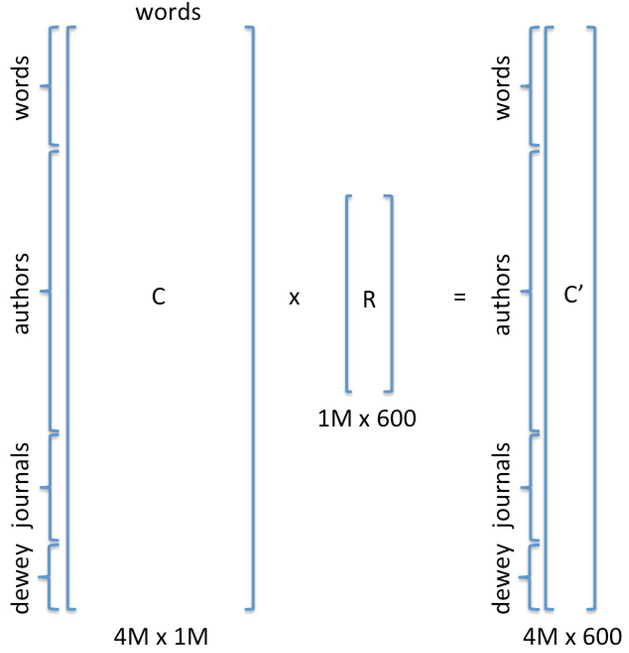}
\caption{Dimensionality reduction using random projection, where $R$ is a random matrix containing $-1, 1$.  \label{fig.rp}}
\end{figure}}

One issue for web-based interactive interfaces is responsiveness. 
This requires scalability of the underlying algorithms. 
For our interface, we solved this problem by combining an \textbf{off-line} preparation phase with an \textbf{on-line} process. Off-line, we build the semantic representation of each entity. Hereby, we use Random Projection to reduce dimensionality (\fig{rp}). In the \textbf{on-line} interface terms from a query are matched to entities in this reduced semantic matrix. The number of hits is further reduced to render a network layout easy to overview and navigate.  

\paragraph{Off-line: Low-dimensional semantic representation using random projection}

1). 

The relatedness of entities is calculated based on the \textit{context} they share, instead of direct co-occurrences in the data. The context is computed as follows. We use a database of documents, in this case journal articles, from which, in this first implementation, we focus on titles and abstracts. After removing stop words,
we select approximately one million most frequent terms (single words or two-word phrases)
as topical terms in this corpus. These terms are used both to represent context (columns of $C$, \fig{rp}) and as searchable topical terms (first rows of $C$), although this is not a requirement. Each entity --- in our case, topical term, author, journal, and Dewey decimal class, as shown in matrix $C$ of \fig{rp} --- is therefore represented by the vector of co-occurrence frequencies between the entity and these terms, accumulated over articles. Authors, for instance, are represented in terms of the vocabulary they use across all their articles. The row vectors define the context for each entity. Cosine similarity calculated from these row vectors defines the semantic similarity between entities. As a consequence, it is possible to compute similarities between entities of different types, such as between a journal and an author, or between a topic and a Dewey class. 

The high dimensionality of the vector representation
is necessary to capture the long tail of word frequencies, but makes direct computation of similarities intractable. To make the algorithm scale, we reduce the dimensionality of the column vectors from 1M to 600 dimensions 
using Random Projection~\cite{Achlioptas2003671,johnson84extensionslipschitz}. The choice of keeping 600 dimensions was decided empirically for our dataset.  

As a result of this approach, computing the low-dimensional representation and the following on-line matching of the search terms becomes very fast. Although this may seem to be a rather drastic approach, our experiments show that meaningful relationships are preserved. It is as if we would ignore some redundancy in the set of terms.
$R$ is a fixed matrix that is not related to $C$ (contrary to other dimensionality reduction methods), so that creating it can be considered an unit-cost operation. 
Moreover, since each entry of $C'$ is a just weighted sum of entries in $C$, instead of updating the term frequencies in $C$, we can adjust the relevant entries of $C'$ directly without ever actually storing $C$. 
In the end, the manageable matrix of $C'$ is stored, with each row representing an entity (either a topical term, an author, a journal or a Dewey decimal class in our case). Cosine similarity (implemented in the interface) or any other type of similarity measure can be computed fast between any pair of entities, which is essential for the interactive exploration. We call $C'$ the \textbf{semantic matrix}.

\paragraph{On-line: Interactive exploration of networked entities}




In the interactive web application --- accessible at \url{http://thoth.pica.nl/relate} --- any exploration starts with a query. The user can copy chunks of text into the search window, from single terms to paragraphs. The process looks up each of the entered terms in the semantic matrix $C'$. It retrieves the corresponding vector representation of the involved entities.
Entities (word, author name or ISSN) which do not have an entry in the semantic matrix are ignored. Cosine similarities are then calculated between this averaged representation of the query and every entity in the matrix. The 500 most related entities are kept.

\begin{figure}
\centering
\includegraphics[width=\linewidth]{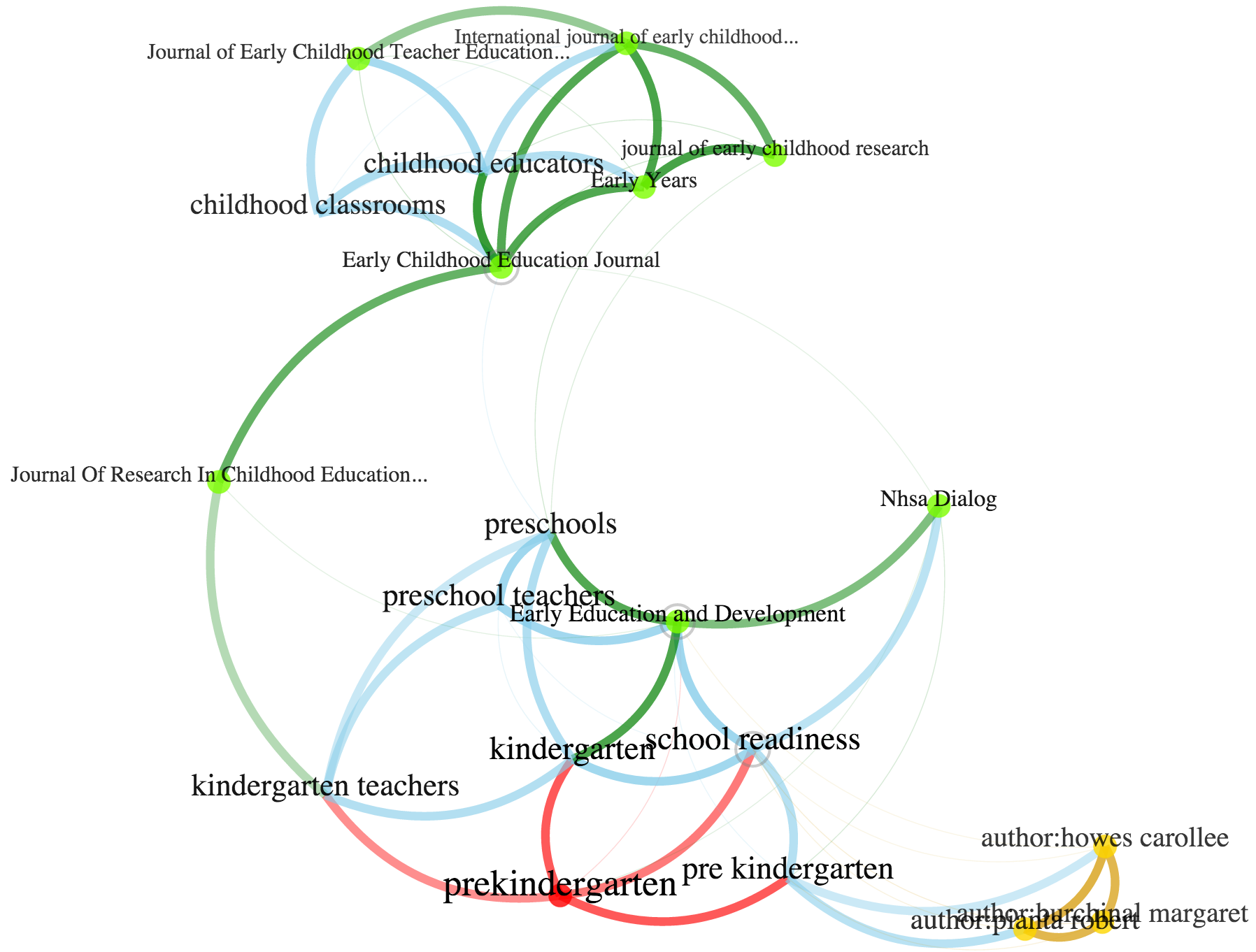}
\caption{Let's start with \textit{prekindergarten} (in red). The 20 most related entities including topics, authors (in yellow) and journals (in green) are presented.  \label{fig.ml}}
\end{figure}

The next step is to filter out non-specific entities which are close to very many other entities. In order to do that, we select 
an entity,  compute the Mahalanobis distance~\cite{mahalanobis1936} between the selected entity and the similarity distribution of its neighbours, and only keep those neighbours that have the smallest Mahalanobis distance. 
The remaining entities are then ready to be projected to a two-dimensional visualisation using multi-dimensional scaling. 
The result is a visualization of a network of related entities (\fig{ml}), of which each node is clickable. Once clicked, a new round of selection starts.

\section{User interface and first observations}

We purposefully kept the interface  plain. Our primary question is: Do the algorithm and resulting network of related entities present a meaningful context to our search quest? Does the network invite to further explore contexts? To enable cross-checking, the upper menu line contains buttons to look up the query in Wikipedia, WorldCat or ScholarGoogle. 


We tested the interface for a couple of different search tasks.\footnote{A first exploration was done by the authors and members of the COST Action TD1210.}
In the \textit{search for an author}, one can detect different spellings in which an author's name occurs in the dataset.\footnote{See the different name variations of \textit{Rienk van Grondel} at \url{http://thoth.pica.nl/relate?input=\%5Bauthor\%3Avan+grondelle+r\%5D}} This could be a first visual screening test to identify issues in author disambiguation or other issues in metadata curation. We also observe that the specificity of a query term influences the search result. 
When \textit{searching for a topic} such as ``machine learning," 
the retrieved context contains related methods such as \textit{neural network}, \textit{svm}, \textit{bayesian classifier} and \textit{decision tree}, as well as related aspects such as \textit{accuracy}, \textit{feature selection}, \textit{overfitting}, and \textit{generalization performance}. The result is not a systematic classification but a very good reflection of the context of this topic in scholarly practice. Sometimes, unexpected terms catch our attention and invite to further investigation. They could also potentially lead to additions to existing classification schemes.
More general or ambiguous terms are also interesting. Searching for ``child care" for instance\footnote{See \url{http://thoth.pica.nl/relate?input=child+care}} reveals many aspects one would expect such as \textit{family income}, \textit{parenting skills}, \textit{child health}, \textit{disadvantaged children}. But we also find \textit{immigrant families}, a term which invites to further exploration.

Because similarities are calculated between all different types of entities simultaneously, it depends very much on the query entry which types show-up in the resulting network. The minimalistic interface allows the user to restrict the display to one type, and this way similarities in only the journal or the author space become visible. The seamless transfer between them allows a first rough delineation of a query in a topical space encompassing terms, authors, journals and Dewey decimal classes.


\section{Future work and discussion}

The interface and its underlying algorithmic processes realise principles of interactive visual browsing and navigating \textit{in vivo} in a large bibliographic database. It builds on past and present attempts to integrate visual elements to on-line searching and browsing \cite{brooks2013,hall2012,kaizer2005,shneiderman2000axes}. 
Pursuing this work we will address the following tasks: (1) We will compare the applied algorithm with other algorithms used in bibliometrics for the delineation of topics and fields. (2) We will pursue more systematic user studies to investigate the role of context visualisation for information foraging. (3) Ultimately, we would like to enable the user to also retrieve related articles. The latter task entails a shift from an explorative interface to an implementation into the service of ArticleFirst. 

There are also a couple of extensions of the method and its application we think are interesting. For example, could we incorporate temporal analysis --- a timeline button? How can the visualisation be improved --- algorithm-wise and design-wise? Could results be exported for secondary analysis? Could we apply sentiment analysis to add connotation to the links? Could elements of recommendations and knowledge discovery be added to the interface and if so in which way?

For the time being, the association we got in our experiments with the interface was to exploring a library through different kinds of catalogues: author catalogue and systematic catalogues. Sometimes, the triangulation between the landing points of search arrows into the space of author, journal, terms and subject headings would lead to an appropriate and comprehensive overview; sometimes it would lead to surprising associations revealing a meaning after closer inspection; and sometimes it would leave us with a labyrinth requiring a thread from Ariadne. It is possible that this is the best which we can expect from a navigable interface into large search spaces: that it complements other search options and enhances those by presenting overview and contextualisation in a tentative way, inviting further exploration, navigation, and eventually close inspection.  

\textbf{Acknowledgement}: Part of this work has been funded by the COST Action TD1210 KnowEscape.

\bibliographystyle{acm-sigchi}
\bibliography{refs}


\end{document}